%% file: main.tex
\documentclass[12pt]{article}

\usepackage[utf8]{inputenc}
\usepackage{wrapfig}
\usepackage{stfloats}

\usepackage{listings}
\usepackage{color,xcolor}
\usepackage{multirow}
\usepackage{sverb}
\usepackage{graphicx}
\usepackage{soul}
\usepackage{xspace}
\usepackage{colortbl}
\usepackage{subfig}
\usepackage[shortlabels]{enumitem}

\usepackage[norndcorners,customcolors]{hf-tikz}
\hfsetbordercolor{yellow}
\hfsetfillcolor{yellow}

\usepackage{times}

\usepackage{verbatim}

\usepackage{tablefootnote}

\usepackage{threeparttable}

\usepackage[utf8]{inputenc}
\usepackage[T1]{fontenc}
\usepackage{lmodern}
\usepackage[ngerman,english]{babel}
\usepackage{amsmath}
\usepackage{mathtools}
\usepackage{graphics}
\definecolor{comment}{cmyk}{0.82,0.39,0,0.25}
\usepackage[prependcaption,textsize=footnotesize,linecolor=comment,backgroundcolor=comment!25,bordercolor=comment]{todonotes}
\usepackage[ruled, vlined, linesnumbered, noresetcount]{algorithm2e}
\usepackage{makecell} %add new line in table cell
\usepackage{pifont}
\newcommand{\cmark}{\text{\ding{51}}}%
\newcommand{\xmark}{\text{\ding{55}}}%
\usepackage{flexisym} %prime symbol
\usepackage{listings}
\usepackage{float}
\usepackage{alltt}

\usepackage[many]{tcolorbox}
\usetikzlibrary{shadows}

\newtcolorbox{shadedbox}{
 colback=red!5!white,
  colframe=red!75!black,
  sharp corners,
  boxrule=0.3mm,
  size=fbox,
  fonttitle=\bfseries
  left=200mm
}
\newcommand{\header}[1]{\textbf{\textsc{#1}}}

\definecolor{mGreen}{rgb}{0,0.6,0}
\definecolor{mGray}{rgb}{0.5,0.5,0.5}
\definecolor{mPurple}{rgb}{0.58,0,0.82}
\definecolor{backgroundColour}{rgb}{0.95,0.95,0.92}

\newcommand{\ma}[1]{\textbf{\color{orange}\emph{M: #1}}}
\newcommand{\machanged}[1]{\color{black}{#1}}

\definecolor{bottlegreen}{rgb}{0.0,0.42,0.31}

\newcommand{\Omit}[1]{}

\colorlet{shadecolor}{gray!40}

\usepackage[normalem]{ulem} %strikeout text
\usepackage{pxfonts} %bold keywords in listing

\usepackage{enumitem}
\setlist{nosep,leftmargin=\parindent}

\newtheorem{definition}{Definition}

\usepackage{flushend}
\usepackage{cleveref}
\usepackage{authblk}

\usepackage{wasysym}
\definecolor{trueColor}{RGB}{98,152,61}
\definecolor{falseColor}{RGB}{192,0,0}

\input{macros}

\begin{document}

\title{
\name: A Technique for Static Configuration-Logic Identification
}
\author[1]{Mohannad Alhanahnah}
\author[2]{Philipp Schubert}
\author[1]{Thomas Reps}
\author[1]{Somesh Jha}
\author[2]{Eric Bodden}
\affil[1]{University of Wisconsin-Madison}
\affil[2]{Paderborn University}
\date{}

\maketitle

\begin{abstract}
\input{abstract}

\end{abstract}

\input{introduction}
\input{background}
\input{manual_study}

\input{combinedProblemStmt}
\input{approachV2}

\input{evaluation}

\input{threatsToValidity}
\input{relatedWork}

\input{conclusion}

\bibliographystyle{plain} 
\bibliography{ref,config,partition,graph,osdi}

\end{document}

%% file: macros.tex
\newcommand{\rqOne}{\textbf{RQ1}}

\newcommand{\rqThree}{\textbf{RQ3}}

\newcommand{\name}[0]{\textsc{slash}\xspace}

\newcommand{\code}[1]{\texttt{#1}}
\newcommand{\lmcas}{\mbox{\textsc{LMCAS}}\xspace}
\newcommand{\neck}{\mbox{\textsc{boundary}}\xspace}
\newcommand{\necks}{\mbox{\textsc{boundaries}}\xspace}
\newcommand{\cli}{\mbox{command-line input}\xspace}
\newcommand{\sota}{\mbox{state-of-the-art}\xspace}
\newcommand{\no}{\textcolor{red}{\xmark}}
\newcommand{\yes}{\textcolor{green}{\cmark}}

\newcommand{\Start}{{v_{{\textit{en}}}}}
\newcommand{\Exit}{{v_{{\textit{ex}}}}}

\newcommand{\host}{{\textit{host}}}

%% file: abstract.tex
Researchers have recently devised tools for debloating software and detecting configuration errors.
Several of these tools rely on the observation that programs are composed of an \textit{initialization phase} followed by a \textit{main-computation phase}.
Users of these tools are required to \emph{manually} annotate the \neck that separates these phases, a task that can be time-consuming and error-prone.
(Typically, the user has to read and understand the source code or trace executions with a debugger.)
Because errors can impair the tools' accuracy and functionality, the manual-annotation requirement hinders the ability to apply the tools on a large scale.

In this paper, we present a field study of 24 widely-used C/C++ programs, identifying common \neck properties in $96\%$ of them.
We then introduce \name, an automated tool that locates the \neck based on the identified properties.
\name successfully identifies the \neck in $87.5\%$ of the studied programs within 8.5 minutes, using up to 4.4 GB memory.
In an independent test, carried out after \name was developed, \name
identified the \neck in $85.7\%$ of a dataset of $21$ popular C/C++ GitHub repositories.
Finally, we demonstrate \name's potential to streamline the \neck-identification process of software-debloating and error-detection tools.

%% file: introduction.tex
% \vspace{-0.25cm}
\section{Introduction} \label{sec:introduction}
Software configurability has emerged as a significant focus in contemporary research ~\cite{Lotrack_TSE, 10.1145/1806799.1806864,Lotrack_ASE, cloudConfig,knobs_FSE15,configuration_icse11,ConfSuggester,configurationAPIs,ConfMapper}.
Concurrently, several initiatives proposed to elevate configurability as a first-class programming element~\cite{ACCORD} and aimed to forge consensus and promote best practices~\cite{ICSE-SEIP20,TSE-SLR}.
One best practice is to organize programs to operate in two phases:
(i) a phase for \emph{initialization}, where configuration logic checks parameters and initializes corresponding values to control the program's activities, and
(ii) a \emph{main-computation phase} that performs actions in accordance with the specified configuration).
One would hope that this structure is reflected in the code---i.e., there is a \neck between the configuration logic and the main computation.
A number of recent papers~\cite{LMCAS,Temporal,PCHECK,Dossier} also describe the advantages that this separation provides for the sake of configuration traceability~\cite{ICSE-SEIP20}, forensic analysis~\cite{Dossier}, optimizing programs~\cite{LMCAS,Temporal}, and detecting configuration errors~\cite{PCHECK}.

PCHECK~\cite{PCHECK} automatically generates a checker that detects configuration errors early to minimize damage from failures. It adds a call to the checker at the end of the initialization phase,
as illustrated in Figure~\ref{fig:pcheck}, which depicts the \neck location in the program \texttt{Squid}, a widely used open-source Web proxy server that supports $237$ configurations.
However, a user of PCHECK needs to \emph{manually} annotate
the \texttt{Squid} source code with the \neck location.

The situation is similar for program-debloating tools.
Temporal-specialization~\cite{Temporal} disables system APIs after the completion of the initialization phase, but also requires the tool user to annotate the \neck.
LMCAS~\cite{LMCAS} specializes programs by executing them up to the \neck to capture the program's state according to the supplied inputs, where again the LMCAS user must annotate the location of the \neck.
      
\noindent
\begin{minipage}{.6\columnwidth} 
  \includegraphics[width=\linewidth]{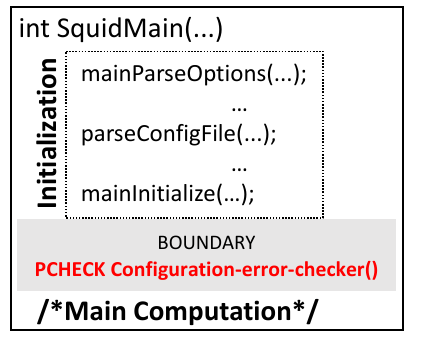}
\end{minipage}%
\begin{minipage}{.4\columnwidth} 
  \captionof{figure}{Location to invoke the configuration-error checker generated by PCHECK~\cite{PCHECK} in Squid---i.e., the \neck at the end of the initialization phase. PCHECK users need to annotate this location \emph{manually}. 
  }
    \label{fig:pcheck}
\end{minipage}

In the absence of a method to assist developers in identifying the \neck, the \textit{manual} annotation is a time-consuming and error-prone task:
the user has to read and understand the source code or trace executions with a debugger.
Because \neck-identification errors can impair the tools' accuracy and functionality, the manual-annotation requirement hinders the ability to apply the tools on a large scale.

We thus first conduct a manual field study to comprehend how the \neck is implemented and to discern its distinguishing characteristics.
Our corpus contains $24$ widely used C/C++ programs (Table~\ref{tab:motivation:prelimStudy}) that employ configuration files or command-line parameters.
Our study identifies various categories of \necks: single-element \necks, multi-element \necks, or ``blended'' (i.e., no \neck).
The study further indicates that 23 (96\%) of the programs possess a single-element \neck.
Accordingly, we developed \name, a tool to identify a \neck automatically.
\textbf{\name focuses on the common case of identifying a \neck in applications that contain one or more single-element \necks.}
\name analyzes LLVM IR, and targets programs written in an imperative programming language, such as C or C++.

Our work makes the following contributions:
\begin{enumerate}
  \item
    We conducted a manual field study to determine (i) to what extent real-world programs contain a \neck that separates configuration logic from the main-computation logic, and (ii) for programs that do contain a \neck, what structural patterns can be used to identify the end of the initialization/configuration phase (\S\ref{sec:manualStudy}).
  \item 
    We present an algorithm that either identifies a \neck that separates the initialization and main-computation phases, or reports that it was unable to do so (\S\ref{sec:approach}). 
  \item
    We implemented the {\neck}-identification algorithm in a tool, called \name,
    and evaluated it on (a) the $24$ programs used in the manual field study,
    and (b) $21$ popular C/C++ Github repositories not part of the manual field study
(\S\ref{sec:evaluation}). 
    
  \item
    We demonstrate that the \necks that \name identifies are suitable substitutes for the ones identified manually (by the respective developers) for two software-debloating tools~\cite{Temporal,LMCAS} and a configuration-error-checking tool \cite{PCHECK}.
    
\end{enumerate}

%% file: background.tex
% \vspace{-0.25cm}
\section{Background} \label{section}
This section provides
background on some concepts and patterns that we relied on in our manual field study. 

\vspace{-0.2cm}
\subsection{Program Phases}
\begin{figure}[tb!]
    \raggedright
    \includegraphics[width=\linewidth, width=8.5cm]{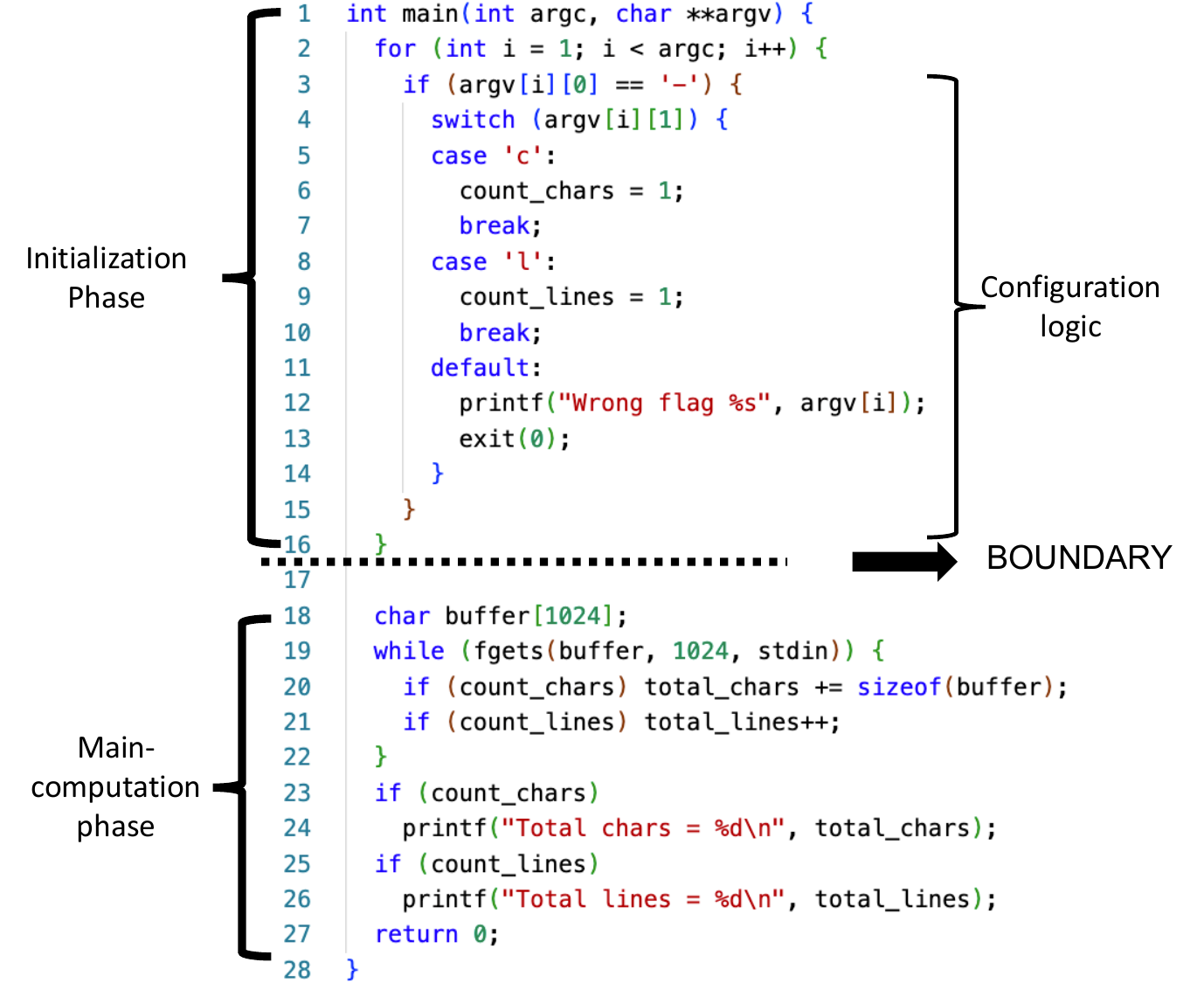}
    \caption{A scaled-down version of the \texttt{wc} utility.
    The \neck 
    % between the initialization phase and main-computation phase 
    could be just before line $17$, just before line $18$, or just before line $19$.
    \lmcas executes \texttt{wc} up to the \neck during the course of its analysis but its users are required to annotate the \neck location \emph{manually}.
}
    \label{fig:illustrativeEx}
\end{figure}

An example of a \neck is shown in Figure~\ref{fig:illustrativeEx}, which represents a scaled-down version of the UNIX word-count utility \texttt{wc}:
\texttt{wc} reads a line from \texttt{stdin}, counts the number of lines and/or characters in the input stream,
and prints the results on \texttt{stdout}.
This program has two phases, the code for which is found in disjoint regions: 
\begin{itemize}
  \item
    The \emph{initialization phase} starts at the entry point of \texttt{main} (line $1$), and ends at line $16$.
  \item
    The \emph{main-computation phase} starts at line $18$ and continues to the end of \texttt{main} (line $28$).
\end{itemize}

\noindent When the configuration logic in the initialization phase is executed, a parameter expressed in some \emph{external format}---here \texttt{argv[1]} as a C string---is translated to an \emph{internal format} and assigned to one or more program variables that host run-time configuration data. 
These variables are known as \emph{configuration-hosting variables}~\cite{trimmer_journal}.
In Figure~\ref{fig:illustrativeEx}, after the configuration logic executes, the configuration-hosting variables \texttt{count_char} and \texttt{count_line} each hold internal-format values of \texttt{0} or \texttt{1}.
The main-computation logic
then performs the primary processing function of the program, with its actions controlled by the values of \texttt{count_char} and \texttt{count_line}.

The two regions are tied together through the values of the configuration-hosting variables:
when the main-computation phase executes, the values of
\texttt{count_char} and \texttt{count_line} control which portions of the main logic execute.
In \texttt{wc}, for instance, \texttt{count_char} controls whether lines $20$ and $24$ execute, and \texttt{count_line} controls whether lines $21$ and $26$ execute. In \texttt{wc}, there is a \neck at line $17$ that physically separates the configuration logic from the main-computation phase.

\begin{comment}
\vspace{-0.2cm}
\subsection{Run-Time Configurations}
Run-time configurations allow a program's behavior to be changed by users post-deployment without re-compiling the program~\cite{TSE-SLR}.
Run-time configurations are typically specified by setting values of configuration parameters in a configuration file, or by passing them in the command line when the program is invoked~\cite{cloudConfig}.
Compared to compile-time configurations controlled by \texttt{\#ifdef}-based feature flags~\cite{ICSE-SEIP20}, run-time-configuration mechanisms use more complex data types (e.g., string and numeric) and have different error patterns.
\end{comment}

\vspace{-0.2cm}
\subsection{Configuration-Logic Phases}
To inform the process of identifying a \neck, it is crucial to understand the configuration logic and patterns inside the initialization phase. We adopt the taxonomy of configuration design
of Zhang et al.~\cite{cloudConfig}, which involves
the following configuration phases:
\begin{enumerate}
    \item \textbf{Parse and Assign:} run-time-configuration information is first parsed and translated. Translated values of configuration parameters are typically Booleans, integers, or strings.
    In a \textbf{command-line program}, the inputs are provided via command-line arguments.
    In C/C++ programs, command-line arguments are passed to \texttt{main()} via the parameters \texttt{argc} and \texttt{argv}: 
    \texttt{argc} holds the number of command-line arguments,
    and \texttt{argv[]} is an array (of that length) of pointers;
    the elements of \texttt{argv[]} point to the different arguments passed to the program.
    The application then assigns values of \texttt{argv[]} elements to configuration-hosting variables according to a predefined argument-value mapping. 
    
    \hspace*{1.5ex}
    Similar logic is used in \textbf{configuration-file programs}. They also permit command-line arguments,
    yet they receive further arguments using configuration files, whose location is typically provided via one command-line argument.
    The configuration file is frequently parsed using a system-call API.
    For example, \textit{Nginx} uses the Linux system call \texttt{pread},
    % \footnote{https://man7.org/linux/man-pages/man2/pwrite.2.html}
    and \textit{DNSProxy} uses the C library function \texttt{fgets}.
    % \footnote{https://man7.org/linux/man-pages/man3/fgets.3p.html}
    The application then usually assigns configuration-hosting variables values according to a predefined keyword-value mapping.
    
    \item \textbf{Check} and \textbf{Exception/Error Handling:} in general, these steps are intertwined with the parse-and-assign step.
    They validate the provided inputs based on certain constraints, identify incorrect configuration parameters and---if present---provide user feedback and terminate the program.
\end{enumerate}

\noindent
% The parse-assign-check steps are executed within a loop until all configuration parameters have been processed (or an error is encountered). If the provided configuration parameters pass the checks, then the main computation of the program can use their (translated) values to select the desired functionalities. The point at which such parameter processing is completed represents the beginning of the program's main-computation logic; i.e., it is the \neck between the initialization phase and the main-computation phase.
Parse-assign-check steps are executed inside a loop until all configuration parameters are processed or an error arises. Once parameters pass checks, the main program can utilize these (translated) values to select functionalities. This processing completion denotes the transition from the initialization to the main-computation phase.

%% file: manual_study.tex
% \vspace{-0.2cm}
\section{Understanding \neck Characteristics} \label{sec:manualStudy}
This section describes our manual field study, conducted to determine (i) to what extent real-world programs contain a \neck that separates configuration logic from the main-computation logic, and (ii) for programs that do contain a \neck, what structural patterns exist that could be used to automate the process of identifying the \neck. 

\input{manual_study_programs.tex}
\vspace{-0.2cm}
\subsection{Methodology}
\noindent
\textbf{Selection of subject programs.}
We manually inspected $24$ widely-used~\cite{LMCAS,Temporal,PhASAR,10.1145/3315568.3329965,ConfMapper} end-user and server C/C++ programs, listed in Table~\ref{tab:motivation:prelimStudy}. 
The configurations of these programs are provided either through command-line arguments or configuration files. 
The end-user programs include utilities (i.e., \texttt{sort}, \texttt{objdump}, \texttt{diff}, and \texttt{gzip}).
The server programs include web servers (i.e., \texttt{Nginx}), DNS servers (i.e., \texttt{DNSProxy}), and database programs (i.e., \texttt{PostgreSQL}).  

\noindent
\textbf{Manual-inspection procedure.} \label{subsec:manualInspection}
We manually inspected the source code of the programs to see if we could identify a \neck location.
The inspection was conducted by one person from our team, but a second opinion was obtained for challenging cases, such as Nginx, httpd, and PostgreSQL.
The manual field study was performed as follows: 
\begin{enumerate}[(a)]
    \item
      We built the program and ran it with \texttt{-help} to display all runtime
      configuration parameters.
      For config-file programs, we also inspected the program's default configuration-file templates to identify the set of predefined keywords (e.g., \texttt{Nginx} uses the directive \textit{gzip} to enable/diable compression).
    \item
      Next, we identified the entry-point function in the source code.
      Because the study considered C/C++ programs, we searched for a function named ``\texttt{main}'' that has two parameters named ``\texttt{argc}'' and ``\texttt{argv}.''
    \item
      \label{It:IdentifyParseAssignCheckLoop}
      We identified the locations in the source code where the %run-time 
      configuration parameters are parsed, assigned, and checked (thereby identifying the configuration-hosting variables).
      A regular-expression search (based on the knowledge gained from step (a) was sufficient for identifying such locations.
      We observed that some programs parse the configuration parameters
      outside of \texttt{main};
      for instance, the \texttt{main} function of a command-line program might invoke another function and pass \texttt{argv} as a parameter.
      For configuration-file programs, we performed a regular-expression search to find (i) method names in APIs for reading/parsing files, and (ii) keywords used in the configuration file. 
      
      \hspace*{1.5ex}
      To supplement source-code inspection, we ran the programs with a debugger (GDB) to track the use of \texttt{argv} and identify the location where the provided configuration parameters are parsed. 
      Similarly, we ran the configuration-file programs with GDB by specifying the configuration-file-parsing API as breakpoints. 
    \item
      \label{It:IdentifyNeckLocation}
        We sought to identify a location for the \neck.
      We looked to see if the location just after the end of the loop containing the parse-assign-check logic was acceptable.
      In some cases, the \neck location was a bit further along in the program because the values of configuration-hosting variables are sometimes set or adjusted after the parse-assign-check loop when one configuration feature overrides another.
\end{enumerate}

% \vspace{-0.3cm}
\subsection{Results} \label{sec:discussionPrelem}
\noindent
\textbf{Categories of \necks.} Our study identifies several types of \necks within programs:
\begin{itemize}
  \item
    \textbf{Single-element \necks}: one or more sites exist that, individually, are each an acceptable \neck location (as in \texttt{wc} from Figure~\ref{fig:illustrativeEx}).
  \item
    \textbf{Multi-element \necks}:
    No single-element \neck exists, but a collection of sites separate the program's configuration logic from the main-computation phase.
  \item
    \textbf{``Blended'' \necks}: the application's configuration logic is ``blended'' into the main-computation phase, yielding no clear \neck of the foregoing two types.
\end{itemize}

\noindent The manual-field study showed that $23$ programs possess single-element \necks.
As illustrated in Figure~\ref{fig:illustrativeEx}, it is possible that there are multiple locations where a single-element {\neck} could be located.
(The term ``multiple single-element \necks'' should not be confused with ``multi-element \neck'').
E.g., in Figure~\ref{fig:illustrativeEx} the {\neck} could be just before line $17$, just before line $18$, or just before line $19$.
Column 4 of Table~\ref{tab:motivation:prelimStudy} indicates which of the programs had multiple single-element {\necks}.
The remaining program, \texttt{PostgreSQL} (a database server), has a multi-element \neck.
It is a Swiss-Army-knife program, consisting of several stand-alone programs, each with its own \neck or \necks.

\begin{table}[]
\centering
\caption{Number of programs in each \neck category.}
\vspace{-0.4cm}
\scalebox{0.9}{
\begin{tabular}{|c|c|c|c|}
\hline
Dataset            & Single-elem. & Multi-elem. & Blended \\ \hline
Manual field study & 23     & 1             & 0       \\ \hline
Previously unseen  & 17     & 1             & 3       \\ \hline         
\end{tabular}
}
\label{tab:neckCategories}
\end{table}

The corpus of programs used in the manual field study (Table~\ref{tab:motivation:prelimStudy}) had no instances of ``blended'' \necks.
However, our evaluation dataset of previously unseen programs (Table~\ref{tab:evaluation:unseen}) contains three programs with ``blended'' \necks. Table~\ref{tab:neckCategories} shows the numbers of programs in each \neck category.

,

\noindent
\textbf{Behaviour of configuration-hosting variables.}
In $20$ of the $24$ programs, the values of the configuration-hosting variables are not changed after they are given values within the parse-assign-check loop. In three cases (\texttt{tcpdump}, \texttt{Gzip}, and \texttt{readelf}), we observed that feature interactions---in one case, involving a compile-time configuration feature (\texttt{\#ifdef})---can cause the program to change the value of a configuration-hosting variable after the parse-assign-check loop.
\begin{shadedbox}
{\bf \header{Field Study Results:}}
The study revealed that:
\begin{itemize}[leftmargin=2.5ex]
  \item
    $23$ of the $24$ programs have a
    single-element \neck that divides the program into its configuration logic and its main-computation logic.
  \item
    The assignments to configuration-hosting variables typically all occur inside a parse-assign-check loop that processes the command line or configuration file.
  \item
    Because of interactions among configuration features, the values of configuration-hosting variables are sometimes adjusted after the parse-assign-check loop.
\end{itemize} 
\end{shadedbox}

\vspace{-0.4cm}
\subsection{Findings: Properties of the {\neck}} \label{sec:stucturalProp}

Our study revealed a set of properties that were common across the $23$ programs in which we were able to identify a single-element \neck.
We use these properties in \S\ref{sec:TheBoundaryIdentificationProblem} and \S\ref{sec:approach} to address the problem of \emph{automatically} identifying
a suitable single-element {\neck} location.

The main property that we can infer from Figure~\ref{fig:illustrativeEx} is that the {\neck} divides a program's control-flow graph into two disjoint subgraphs.
This property implies that the {\neck} is a so-called \textit{articulation point} in the program
(i.e., a vertex cut of the control-flow graph, of size $1$).
In addition, we observed that the {\neck} should be:
\begin{itemize}
  \item
    \textbf{Reachable} from the entry point of the program.
  \item
    \textbf{Executed} exactly once, which eliminates the possibility that the {\neck} is inside a loop or a conditional statement. 
\end{itemize}

\noindent
To  validate our hypothesis,
we instrumented the programs by adding a print statement at the location where the \neck is identified (e.g., line $17$ in Figure~\ref{fig:illustrativeEx}) and then executed the program several times with different sets of %run-time-
configuration parameters.
If the print statement is executed, the run indicates the \neck is reachable from the entry;
if the print statement is printed only once, the run indicates that the \neck is executed exactly once (for the given %run-time 
configuration).

In the debugging performed in step \ref{It:IdentifyParseAssignCheckLoop} of the manual inspection steps in \S~\ref{subsec:manualInspection}, we dynamically traced the uses of \texttt{argv} and calls to configuration-file parsing APIs as a way to pinpoint the locations where %run-time-
configurations are parsed, assigned, and checked.
This use of dynamic tracing of \texttt{argv} during manual analysis suggests that a technique to identify {\necks} automatically will need to perform dataflow analysis to track uses of \texttt{argv} statically.
\vspace{-0.2cm}
\begin{shadedbox}
%\hspace{-0.5cm}
{\bf \header{Findings:}}
Properties relevant to identifying automatically a program's \neck include:
\begin{enumerate}[leftmargin=3.5ex]
  \item
    Configuration-hosting variables are data dependent or control dependent on \texttt{argv}.
  \item
    The \neck should be located after at least one loop.
  \item
    The \neck represents an articulation point in the program's control-flow graph.
  \item
    The \neck should be reachable from the entry point and executed only once.
\end{enumerate}
\end{shadedbox}

%% file: manual_study_programs.tex
\begin{table}[tb!]
\centering
\caption{
Results of the manual field study.
Column $2$ gives LOC (in thousands),
based on readable LLVM IR.
Column $3$ indicates whether the program can receive additional configuration settings through a configuration file.
Column $4$ indicates if there are multiple acceptable (single-element)
\neck locations. Column $5$ specifies whether the location selected based on our manual inspection
(typically the first acceptable \neck location)
lies within or outside \texttt{main}.
}
\scalebox{0.75}{
\begin{tabular}{|c|r|c|c|c|}
\hline
\multicolumn{1}{|c|}{\textbf{Program}} & {\textbf{kLOC}} & {\textbf{{Config File}}} & {\textbf{\shortstack{Multiple \neck}}} & \textbf{Inside main} \\ \hline
\rowcolor{gray!40} \multicolumn{5}{|c|}{End-User Programs}  \\ \hline                      
\multicolumn{1}{|c|}{curl-7.47.0}      & 31.6           & {\yes}                             & {\yes}                                      & \no                     \\ \hline
\multicolumn{1}{|c|}{date-8.32}      & 56.9           & {\no}                             & {\no}                                      & \yes                     \\ \hline
\multicolumn{1}{|c|}{diff-2.8}         & 37.2           & {\no}                             & {\no}                                      & \yes                     \\ \hline
\multicolumn{1}{|c|}{du-8.32}        & 109.1         & {\no}                             & {\no}                                      & \yes                     \\ \hline
\multicolumn{1}{|c|}{echo-8.32}      & 11.0           & {\no}                             & {\no}                                      & \yes                     \\ \hline
\multicolumn{1}{|c|}{gzip-1.2.4}       & 26.7           & {\no}                             & {\yes}                                      & \yes                     \\ \hline
\multicolumn{1}{|c|}{id-8.32}        & 15.1           & {\no}                             & {\no}                                      & \yes                     \\ \hline
\multicolumn{1}{|c|}{kill-8.32}      & 12.1           & {\no}                             & {\no}                                      & \yes                     \\ \hline
\multicolumn{1}{|c|}{objdump-2.33}     & 1049.0        & {\no}                             & {\no}                                      & \yes                     \\ \hline
\multicolumn{1}{|c|}{psql-15}     & 189.1        & {\no}                             & {\yes}                                      & \yes                     \\ \hline
\multicolumn{1}{|c|}{readelf-2.33}     & 413.7         & {\no}                             & {\yes}                                      & \yes                     \\ \hline
\multicolumn{1}{|c|}{sort-8.32}      & 55.9           & {\no}                             & {\no}                                      & \yes                     \\ \hline
\multicolumn{1}{|c|}{tcpdump-4.10.0}   & 608.6         & {\no}                             & {\no}                                      & \yes                     \\ \hline
\multicolumn{1}{|c|}{uniq-8.32}      & 14.8          & {\no}                             & {\no}                                      & \yes                     \\ \hline
\multicolumn{1}{|c|}{wc-8.32}        & 17.8           & {\no}                             & {\yes}                                      & \yes                     \\ \hline
\multicolumn{1}{|c|}{wget-1.17.1}      & 165.5         & {\yes}                             & {\yes}                                      & \yes                     \\ \hline
\rowcolor{gray!40} \multicolumn{5}{|c|}{Server Programs}             \\ \hline
\multicolumn{1}{|c|}{bind-9.15.8}      & 1755.1       & {\yes}                             & {\yes}                                      & \yes                     \\ \hline
\multicolumn{1}{|c|}{DNSProxy-1.17}    & 3.4            & {\yes}                             & {\yes}                                      & \yes                     \\ \hline
\multicolumn{1}{|c|}{httpd-2.4.51}     & 179.0         & {\yes}                             & {\yes}                                      & \yes                     \\ \hline
\multicolumn{1}{|c|}{knockd-0.5}       & 10.9           & {\yes}                             & {\yes}                                      & \yes                     \\ \hline
\multicolumn{1}{|c|}{lighttpd-1.4.54}  & 174.0         & {\yes}                             & {\no}                                      & \no                     \\ \hline
\multicolumn{1}{|c|}{mini-httpd1-1.19} & 16.4          & {\yes}                             & {\no}                                      & \yes                     \\ \hline
\multicolumn{1}{|c|}{Nginx-1.19.0}     & 589.2         & {\yes}                             & {\yes}                                      & \yes                     \\ \hline
\multicolumn{1}{|c|}{PostgreSQL-15} & 4626.3         & {\yes}    & \multicolumn{2}{c|}{Multi-element \neck}     \\ \hline
\end{tabular}
}
\label{tab:motivation:prelimStudy}
\end{table}

%% file: combinedProblemStmt.tex
\section{The \neck-Identification Problem}
\label{sec:TheBoundaryIdentificationProblem}

This section provides an abstract overview of the \neck concept, presenting the process of \neck identification as a form of staging transformation (\S\ref{sec:AbstractView}).
It then defines the \neck-identification task (\S\ref{sec:notation} and \S\ref{sec:problemStmt}).

\subsection{An Idealized View}
\label{sec:AbstractView}

\noindent
At an abstract level, automated {\neck} identification can be thought of as a kind of \emph{staging transformation}~\cite{staging} that isolates (or ``stages'')
the processing of
a program's configuration parameters.
Staging transformations were originally proposed to separate a program's computation into stages for optimization purposes. 
In our context,
given a program $P(x)$ with body $f(x)$, where $x$ represents some configuration parameter, we wish to consider $P$ as having the form shown below in the second line:
\vspace{-0.4cm}
\begin{equation}
  \label{eq:stagingOneParameter}
  \begin{array}{rr@{\hspace{1.0ex}}c@{\hspace{1.0ex}}l}
                & P(x) & = & f(x) \\
   \rightarrow  & P(x) & = & \textbf{let}\; t = \textit{translate}(x)\; \textbf{in} \; g(t)
  \end{array}
\end{equation}
\vspace{-0.4cm}

Here, $\textit{translate}(\cdot)$ converts from the external configuration-specification format to the internal format, and $t$ is a configuration-hosting variable.
Thus,
\begin{itemize}
  \item
    ``$\textbf{let}\; t = \textit{translate}(x)\; \textbf{in}\; \ldots$'' represents the configuration logic of $P$.
  \item
    ``$g(t)$" represents the main logic, which performs the primary processing function of $P$, based on the value of configuration-hosting variable $t$. 
\end{itemize}

\noindent The {\neck}-identification challenge is to find the code that constitutes $\textit{translate}$ within function definition $f$.

This abstract characterization of {\neck} identification permits giving a ``rational reconstruction'' of some previous work.
For instance, both PCHECK~\cite{PCHECK} and Zhang et al.\ \cite{cloudConfig} propose methodologies to ensure that the value of a configuration parameter is checked against an appropriate constraint $\varphi(\cdot)$ on the parameter before it is used.
Thus, if one has program $P(x)$ in the form shown on the second line of Eqn.~(\ref{eq:stagingOneParameter}), the essence of both PCHECK and the Zhang et al.\ paper is to transform $P$ as follows:
\vspace{-0.13cm}
\begin{equation} \label{eq:configCheck}
  \begin{array}{rr@{\hspace{1.0ex}}c@{\hspace{1.0ex}}l}
                & P(x) & = & \textbf{let}\; t = \textit{translate}(x)\; \textbf{in} \; g(t) \\ 
    \rightarrow & P(x) & = & \textbf{let}\; t = \textit{translate}(x)\; \\
                &        &   & \textbf{in} \; \textbf{if}\; \varphi(t)\; \textbf{then}\; g(t)\; \textbf{else}\; \textbf{abort}
  \end{array}
\end{equation}
\vspace{-0.2cm}

Furthermore, several issues discussed in~\cite{ICSE-SEIP20, TSE-SLR,Lotrack_TSE,Lotrack_ASE} can be characterized as ``it is advantageous to separate the configuration logic from the program's main-computation logic for the sake of facilitating configuration tracking and analysis,'' which at an abstract level amounts to:

\vspace{-0.2cm}
\begin{equation} \label{eq:configChar}
  \begin{array}{l}
  \textit{Given program $P$ in the form}
  \\
   \qquad P(x) = \textbf{let}\; t = \textit{translate}(x)\; \textbf{in} \; g(t) \\ 
    \textit{Analyze the usage of}\, t\; \textit{in}\; g(t)  
  \end{array}
\end{equation}
\vspace{-0.2cm}

\noindent
For instance, LOTRACK's usage analysis aims to identify code fragments corresponding to load-time configurations~\cite{Lotrack_TSE,Lotrack_ASE} in Android and Java programs. 

Finally, LMCAS~\cite{LMCAS} relies on manual techniques to identify a program's {\neck}, and then performs partial evaluation \cite{partialEvaluation} with respect to the values of configuration-hosting variables.
Abstractly, LMCAS operates similarly to what was discussed above, except that $P$ is now a two-argument program, $P(x,y)$.
\vspace{-0.1cm}
\begin{equation} \label{eq:lmcas}
  \begin{array}{rr@{\hspace{1.0ex}}c@{\hspace{1.0ex}}l}
                & P(x,y) & = & \textbf{let}\; t = \textit{translate}(x)\; \textbf{in} \; g(t,y) \\ 
    \rightarrow & P_x(y) & = & g_t(y) 
  \end{array}
\end{equation}
\vspace{-0.3cm}

Program $P_x(y)$ is a version of $P(x,y)$ specialized with respect to a specific value of $x$.
The body of $P_x(y)$ is obtained by finding and evaluating $t = \textit{translate}(x)$, and then running a partial evaluator on $g$ with static input $t$ to create $g_t(y)$, which is a version of $g(t,y)$ specialized on the value of $t$.

\vspace{-0.25cm}
\subsection{Terminology and Notation}
\label{sec:notation}

Because the \neck constitutes an articulation point in the program, we formulate the \neck-identification task as a vertex-cut graph-partitioning problem.

\begin{definition}\label{De:ICFG}
Let $G=(V,E,\Start,\Exit)$ denotes the Interprocedural Control-Flow Graph (ICFG) of a program $P$.
Vertices $\Start$ and $\Exit$ represent the entry vertex corresponding to the \texttt{main} and end of the program, respectively.
\end{definition}

Without loss of generality, we assume that each vertex in $G$ is reachable from $\Start$ along a path in which each procedure-return edge is matched with its closest preceding unmatched procedure-call edge (a so-called ``interprocedurally valid path''
\cite[\S7-3]{Chapter:SP81}). A vertex $v$ in the control-flow graph $G_Q$ of some procedure $Q$ is said to be an \emph{articulation point} if removing $v$, and all control-flow edges into and out of $v$, partitions $G_Q$ into two non-empty subgraphs.

\vspace{-0.2cm}
\subsection{Problem Definition} \label{sec:problemStmt}

In the abstract view of the \neck-identification problem discussed in \S\ref{sec:AbstractView}, a command-line or configuration-file program $P(x) = f(x)$ has the form shown on the second line of 
Eqn.\ (\ref{eq:stagingOneParameter}).

Eqn.\ (\ref{eq:stagingOneParameter}) is stated in an abstract form, as if we were considering a program in a functional programming language.
However, we need to translate this idea to something that is suitable for an imperative programming language, such as C/C++.
In such a case, configuration parameter $x$ will be \texttt{argv}.

Our goal is to identify $\textit{translate}(x)$, whose end is considered to be the \neck (but in an imperative program), which leaves us with two questions:

\begin{enumerate}
  \item
    \label{It:ChokePoint}
    What is a suitable ``choke point'' in the program, analogous to the hand-off from ``$t = \textit{translate}(x)$'' to $g(t)$ in Eqn.\ (\ref{eq:stagingOneParameter})? 
  \item
    \label{It:EndOfTranslate}
    What does ``the program has finished $\textit{translate}(x)$'' mean?
\end{enumerate}

With respect to question (\ref{It:ChokePoint}), a natural approach is in terms of the articulation points of the program's control-flow graph $G$:
the candidate choke points are the articulation points of $G$ (denoted by $V_{AP}$).
In general, $G$ can have many articulation points.
We need some other conditions to specify which member of $V_{AP}$ we want:
the \neck separates the vertices of $G$ into the configuration logic (denoted by $V_c$)---which is analogous to $\textit{translate}(x)$---and vertices belong to the main-computation logic (denoted by $V_m$)---which is analogous to $g(t)$.

With respect to question (\ref{It:EndOfTranslate}), discovering the end of $\textit{translate}(x)$ entails identifying the \emph{configuration-hosting variables}~\cite{trimmer_journal}
(which are analogous to variable $t$ in Eqn.\ (\ref{eq:stagingOneParameter})).
These variables are either
(a) assigned %run-time-
configuration values directly, or
(b) control dependent on branch expressions involving %run-time-
configuration quantities.
As supported by the findings from our manual study (\S\ref{sec:manualStudy}), the assignments to configuration-hosting variables
(i) typically all occur inside a parse-assign-check loop that processes the command line or configuration file, but
(ii) some additional assignments to them may occur after the parse-assign-check loop because of interactions among configuration features.
Configuration-hosting variables are typically live variables in the main logic;
moreover, they are used without their values being modified in the main-computation phase~\cite{Lotrack_TSE,CIA}.
We formalize this concept as follows:

\begin{shadedbox}
{\bf \header{Configuration-hosting variables ($C_\host$).}}
\begin{enumerate}[leftmargin=3.5ex]
  \item
    \label{It:ConfigParameter}
    Let $v_x$ denote the CFG vertex that models the passing of configuration parameter $x$ to \texttt{main}.
  \item
    \label{It:VHost}
    Let $V_\host = \{v_0, v_1, \ldots, v_n\}$ denote the set of vertices that represent assignments to configuration-hosting variables:
    $v_i \in V_\host$ if
    \begin{enumerate}[leftmargin=2.5ex]
      \item
        \label{It:VHost:FlowDependent}
        $v_i$ is flow dependent on $v_x$, denoted by $v_x \longrightarrow_f v_i$, or
      \item
        \label{It:VHost:ControlDependent}
        $v_i$ is control dependent on a vertex $w_x$ that uses $x$.
    \end{enumerate}
  \item
    \label{It:CHost}
    The set of configuration-hosting variables $C_\host$ is the set of variables that are assigned to at some member of $V_\host$.
\end{enumerate}
\end{shadedbox}

\noindent
For instance, in the scaled-down word-count program in Figure~\ref{fig:illustrativeEx}, the global variables \texttt{count\_chars} and \texttt{count\_lines} are assigned values at lines $6$ and $9$, respectively;
these assignments are control-dependent on vertices that use \texttt{argv} (i.e., the branch-conditions on lines $3$ and $4$, which play the role of $w_x$ in item (\ref{It:VHost:ControlDependent})).
Thus, by item (\ref{It:VHost:ControlDependent}), $V_\host$ consists of the assignments at lines $6$ and $9$, and
by item (\ref{It:CHost}), $C_\host$ is $\{ \texttt{count\_chars}, \texttt{count\_lines} \}$.

With the concept of $C_\host$ in hand, we can now state the \neck-identification problem.

\begin{shadedbox}
%\hspace{-0.5cm}
  {\bf \header{Problem Definition:}}
  Find an articulation point $B$
  of CFG $G$ that is reachable from $\Start$, and
  \begin{enumerate}[leftmargin=3.5ex]
    \item
        is located after a loop,
    \item
      post-dominates every assignment to a member of $C_\host$, and \label{problemConditionC1}
    \item
      for each $c \in C_\host$, all paths from $B$ to $\Exit$ are free of definitions to $c$. \label{problemConditionC2}
  \end{enumerate}
  Return the closest $B$ to the entry as the \neck.
\end{shadedbox}

The control-flow vertex %at the end of the for-loop 
for line $16$ of Figure~\ref{fig:illustrativeEx}
is an articulation point that meets the conditions of the problem definition:
\begin{itemize}
  \item 
  it is located after the end of the for-loop on lines $2$-$16$
  \item
    it post-dominates every assignment to \texttt{count\_chars} and \texttt{count\_lines}, and
  \item
    all paths from that point to the end of the program are free of definitions to \texttt{count\_chars} and \texttt{count\_lines}.
\end{itemize}
Finally, the articulation point at line $16$ is the closest to the entry point in terms of distance along control-flow edges. 

%% file: approachV2.tex
\section{Automatic \neck Identification}
\label{sec:approach}

\input{algorithm.tex}

This section presents our algorithm to solve the {\neck}-identification problem defined in \S\ref{sec:problemStmt}.
The algorithm is given as Alg.~\ref{alg:boundary-algorithm}.
The discussion of Alg.~\ref{alg:boundary-algorithm}
is structured in two parts, which correspond to lines \ref{alg:ln:BeginAlgorithm}--\ref{alg:ln:bc-taint-two} and \ref{alg:ln:BeginPartTwo}--\ref{alg:ln:return_neck}, respectively.

\begin{itemize}
    \item
    \textbf{Identification of \neck Candidates
    (\S\ref{sec:bc}).}
    This phase identifies a set of \neck candidates, which are a subset of the set of articulation points of the control-flow graph.

    \item \textbf{\neck Identification (\S\ref{sec:filter}).} 
    This phase eliminates all \neck candidates that do not satisfy the three properties of a \neck given in the problem definition in \S\ref{sec:problemStmt}.
\end{itemize}

\subsection{\neck Candidates (Lines \ref{alg:ln:BeginAlgorithm}--\ref{alg:ln:bc-taint-two})} \label{sec:bc}
This phase performs taint analysis and control-flow analysis to identify a set of \neck candidates.
It performs taint analysis to identify the set of configuration-hosting variables $C_\host$ and the set of assignments $V_\host$ defined in \S\ref{sec:problemStmt} (lines \ref{alg:ln:taint_analysis}--\ref{alg:ln:identifyVariablesInAssignments}).
Some adjustments are made when the latter assignments are either inside a loop (lines \ref{alg:ln:computeStronglyConnectedComponents}--\ref{alg:ln:replaceLoopBlocks_end}), or in a procedure other than \texttt{main} (lines \ref{alg:ln:intra_reduction_st}--\ref{alg:ln:reachabilityAndProxies_end}).
In particular, for each candidate not in \texttt{main}, a proxy location is considered just after
the appropriate call in \texttt{main} that would reach the candidate.

The algorithm performs control-flow analysis to identify the set of articulation points in $G$ (line \ref{alg:ln:computeArticulationPoints}).
The outcome of this phase is the intersection between the set of articulation points and the (adjusted) set of vertices that represent assignments to configuration-hosting variables (first performed at line \ref{alg:ln:bc-taint} to reduce the cost of the proxy-finding loop, then at line \ref{alg:ln:bc-taint-two}).

\subsection{\neck Identification (Lines \ref{alg:ln:BeginPartTwo}--\ref{alg:ln:return_neck})} \label{sec:filter}

This phase eliminates all \neck candidates that do not satisfy the three conditions of a \neck given in the problem definition in \S\ref{sec:problemStmt}, namely, each $C \in BC$ must
(C1) be located after a loop,
(C2) post-dominate every assignment to a member of $C_\host$, and
(C3) for each $\textit{var} \in C_\host$, all paths from $B$ to $\Exit$ are free of definitions to $\textit{var}$.
The algorithm removes from $BC$ any vertex $C$ that fails to satisfy all three conditions;
see lines \ref{alg:ln:followsLoop_start}--\ref{alg:ln:followsLoop_end},
\ref{alg:ln:postDominatesLoop_start}--\ref{alg:ln:ft_end}, and
\ref{alg:ln:constant_propagation_start}--\ref{alg:ln:constant_propagation_end}, respectively.

To verify condition (C1), the algorithm checks whether candidate $C \in BC$ is located after a loop (lines \ref{alg:ln:followsLoop_start}-\ref{alg:ln:followsLoop_end}).
To verify condition (C3), the algorithm can use standard techniques---e.g., IFDS-based~\cite{Reps:1995:PID:199448.199462}, inter-procedural constant propagation for each configuration-hosting variable $\textit{var}$, starting at candidate $C \in BC$.
If the analysis reports that $\textit{var}$ is not constant at exit point $\Exit$, then $\textit{var}$ might be (re)defined on some path from $C$ to $\Exit$, and hence condition (C3) is violated.

This phase returns \code{null} if all \neck candidates are eliminated (lines \ref{alg:ln:null_st}--\ref{alg:ln:null_end});
otherwise, it returns the candidate that is closest to $\Start$, the entry point of procedure \texttt{main} (lines \ref{alg:ln:ClosestToEntry} and \ref{alg:ln:return_neck}).
The distance metric used to calculate the closest \neck candidate is the shortest path in terms of control-flow edges.

\subsection{Discussion}\label{sec:algDiscussion}

\textbf{Limitations.}
Alg.~\ref{alg:boundary-algorithm} gives an idealized algorithm that works on an ICFG, yet \name operates on the ICFG only partially.
Specifically, for (a) finding post-dominators,
and (b) finding articulation points our implementation works on individual CFGs in a procedure-by-procedure manner.
Listing~\ref{lst:intra} sketches this variant:
\vspace{-0.14cm}
\begin{lstlisting}[tabsize=2, numberstyle=\tiny, basicstyle=\footnotesize, mathescape, caption={Variant of 
$\textit{computePostDominators}$ and $\textit{computeArticulationPoints}$ used in our implementation of Algorithm~\ref{alg:boundary-algorithm}}, keywordstyle=\color{blue}\bfseries,
morekeywords={foreach},captionpos=b,breaklines=true,numbersep=3pt, frame=tb, label=lst:intra]
$AP,~DTs := \emptyset;$
foreach procedure P { 
        $DTs := DTs~~\cup$ computePostDominators(CFG(P))
        $AP := AP~~\cup$ computeArticulationPoints(CFG(P)) }
\end{lstlisting}
\vspace{-0.14cm}

Consequently, when a \neck candidate is located in some procedure \texttt{p} other than \texttt{main}, the candidate is relocated to the CFG of \texttt{main} by finding a proxy location for the candidate.
We use the basic block of \texttt{main} that contains the call site that calls \texttt{p} (directly or transitively). 
% This approach side-steps the complication of defining a suitable notion of articulation points for ICFGs. 

A second limitation of \name is that it targets only single-element \necks.
When run on a multi-element-\neck program, it could either return the empty set (a correct answer with respect to the question of whether a single-element \neck exists) or some singleton set (which is a false positive).
(In the latter case, our experience is that \name
returns one of the elements of the multi-element 
\neck, and the rest of the elements are other \neck candidates.)
When run on a program in the ``blended''-\neck case, \name returns the empty set (correct with respect to the question of whether a single-element \neck exists).
In this case, it never returns a singleton set because the intertwining of the configuration logic and the main-computation logic causes the properties required of a single-element \neck to be violated.

\noindent
\textbf{Time Complexity.}
The overall worst-case running time of the algorithm is bounded by $\mathcal{O}(|E|\cdot|D|^3 + |N|^2 + |E|)$, where $N$ and $E$ are the sets of nodes and edges, respectively, of the program's ICFG, and $D$ is the domain(s) used in the data-flow problems (taint analysis and constant propagation)~\cite{Reps:1995:PID:199448.199462}.

%% file: algorithm.tex
\begin{algorithm}[H]

% \scriptsize
\tiny
\setlength{\baselineskip}{0.9\baselineskip}

\SetKwFunction{closestReachableFromEntry}{closestReachableFromEntry}
\SetKwFunction{isReachable}{isReachable}
\SetKwFunction{isTransitivelyReachable}{isTransitivelyReachable}
\SetKwFunction{isConstant}{isConstant}
\SetKwFunction{followsLoop}{followsLoop}
\SetKwFunction{isExecutedOnce}{isExecutedOnce}
\SetKwFunction{isReachableFromEntry}{isReachableFromEntry}
\SetKwFunction{isInLoopStructure}{isInLoopStructure}
\SetKwFunction{getLoopExitBlocks}{getLoopExitBlocks}
\SetKwFunction{isInLoopStructure}{isInLoopStructure}
\SetKwFunction{postDominates}{postDominates}
\SetKwFunction{dominatesSuccessors}{dominatesSuccessors}
\SetKwFunction{computeDominatorTrees}{computeDominatorTrees}
\SetKwFunction{computeDominatorGraph}{computeDominatorGraph}
\SetKwFunction{computePostDominators}{computePostDominators}
\SetKwFunction{computeArticulationPoints}{computeArticulationPoints}
\SetKwFunction{computeConstantPropagation}{computeConstantPropagation}
\SetKwFunction{getConfigVars}{getConfigVars}
\SetKwFunction{isStore}{isStore}
\SetKwFunction{isAssignmentInst}{isAssignmentInst}
\SetKwFunction{isConditionalBranchInst}{isConditionalBranchInst}
\SetKwFunction{getLastInst}{getLastInst}
\SetKwFunction{dominatesSuccessors}{dominatesSuccessors}
\SetKwFunction{getAllConfigurationHostingVars}{getAllConfigurationHostingVars}
\SetKwFunction{getBasicBlockOf}{getBasicBlockOf}
\SetKwFunction{identifyVariablesAssignedToIn}{identifyVariablesAssignedToIn}
\SetKwFunction{computeStronglyConnectedComponents}{computeStronglyConnectedComponents}
\SetKwFunction{computeICFG}{computeICFG}
\SetKwFunction{isCondition}{isCondition}
\SetKwFunction{getAssignmentsIn}{getAssignmentsIn}
\SetKwFunction{getEntryPointBasicBlock}{getEntryPointBasicBlock}
\SetKwFunction{isControlDependent}{isControlDependent}
\caption{Single-element {\neck}-identification algorithm.
}
\label{alg:boundary-algorithm}
\DontPrintSemicolon
\KwIn{Program P, SrcProcedure}%\neck properties
\KwOut{\neck}
    % \textcolor{blue}{\tcc{Perform Control-Flow Analysis}}
    % \textcolor{blue}{\tcc{Perform Taint Analysis}}
    % \textcolor{blue}{\tcc{Perform Path Analysis}}
    Entry point $Entry$ = \getEntryPointBasicBlock($P$)\;  
    G = \computeICFG($P$)\;
    Initalize ConfigHostVars $V_\host$ ,BoundaryCandidates $BC$\;
    \textcolor{red}{\tcc{\textbf{Identification of \neck Candidates (Section~\ref{sec:bc})}}}  \label{alg:ln:BeginAlgorithm}
    T = computeTaintAnalysis(G, SrcProcedure);\;  \label{alg:ln:taint_analysis}
	\ForEach{$node \in G$}{ \label{alg:ln:taint_loop_start}
	     \ForEach{$Res \in T.getTaintResultsAt(node)$} { \label{alg:ln:getTaintResultsAt}
	         \ForEach{$Op \in node.operands()$} {
	             \If{$Op$ = $Res~\land$ \isAssignmentInst{$node$}}{
	                 $V_\host\ {\cup}{=}~~ node.getBasicBlock()$\; \label{alg:ln:collectBasicBlocksDirectFlow}
	             }
	           %  \If{$Op$ = $Res~\land$ \isCondition{$Op$} $\land$ \isConditionalBranchInst{$node$}}{
	           %     \ForEach{$s \in node.successors()$}{
    	       %         $V_\host\ {\cup}{=}~$ \getAllAssignmentsIn{$s$}\; \label{alg:ln:collectBasicBlocksIndirectFlow}
    	       %      }
	           %  }
	             \If{$Op$ = $Res~\land$ \isControlDependent{$node$, $Op$}}{
    	                $V_\host\ {\cup}{=}~$ \getAssignmentsIn{$node.getBasicBlock()$}\; \label{alg:ln:collectBasicBlocksIndirectFlow}
	             }
	         }
	     }
    } \label{alg:ln:taint_loop_end}
    $C_\host =$ \identifyVariablesAssignedToIn{$V_\host$}\; \label{alg:ln:identifyVariablesInAssignments}
    \textcolor{blue}{\tcc{Add a loop's successors, if $C \in V_\host$ is within a loop}} %\ma{what is the content of SCCs and how it's used at lines 14 and 31?}}
        $SCCs =$ \computeStronglyConnectedComponents{$G$}\; \label{alg:ln:computeStronglyConnectedComponents}
	\ForEach{$C \in V_\host$}{ \label{alg:ln:replaceLoopBlocks_st}
        \If{\isInLoopStructure{$C, SCCs$}}{
		    $BC~{\cup}{=}$ \getLoopExitBlocks{$C$}\;
	    }\lElse{
	        $BC~{\cup}{=}~C$
	    }
	} \label{alg:ln:replaceLoopBlocks_end}
    $AP = \computeArticulationPoints{G}$\; \label{alg:ln:computeArticulationPoints}
    $BC = BC \cap AP$\; \label{alg:ln:bc-taint}
    \ForEach{$C \in BC$}{ \label{alg:ln:intra_reduction_st} \label{alg:ln:reachabilityAndProxies_start}
        \If{\isReachableFromEntry{$C$}}{ \label{alg:ln:reachable}
            \textcolor{blue}{\tcc{Find proxies in main for \neck candidates in other procedures}}
            \If{$C \notin \texttt{main}$}{ \label{alg:ln:intra_reduction_start}
                \ForEach{$\textit{CallSite} \in \texttt{main}$}{
                    \If{\isReachable{$\textit{CallSite}, C$}}{
                        $BC = BC \setminus \{C\}$\;
                        $BC~{\cup}{=}~ CallSite.getBasicBlock()$\;
                    }
                }
            } \label{alg:ln:intra_reduction_end}
	    }\lElse{$BC = BC \setminus \{C\}$\;}
	} \label{alg:ln:reachabilityAndProxies_end}
	$BC = BC \cap AP$ \textcolor{blue}{\tcc*{Proxies must be articulation points}} \label{alg:ln:bc-taint-two}
    \textcolor{red}{\tcc{\textbf{\neck Identification (Section~\ref{sec:filter})}}} \label{alg:ln:BeginPartTwo}
    $DG =$ \computePostDominators{$G$}\; \label{alg:ln:computeDominators}
    \ForEach{$C \in BC$}{ \label{alg:ln:ft_st}
        \textcolor{blue}{\tcc{C1: Check that \neck is located after a loop}}
        \If{!\followsLoop{$C$}} { \label{alg:ln:followsLoop_start}
		    $BC = BC \setminus \{C\}$\;
            % break\;
		} \label{alg:ln:followsLoop_end}
        \textcolor{blue}{\tcc{C2: Check whether $C$ post-dominates every assignment $d$ to a var in $C_\host$}}
        % \tcc{\twr{The previous loop condition ``for each $v \in V_\host$'' was incorrect: We need to check whether $C$ post-dominates each assignment $d$ to a member of $C_\host$.  Does the implementation do so?}}
        % I think the previous loop condition was also valid, since the taint analysis collects assignment statement, which is why iterating V_{host} should be fine, too.
        $\textit{PostDominatesFlag} = true$\; \label{alg:ln:postDominatesLoop_start}
        \ForEach{assignment $d$ to a variable in $C_\host$}{
            \If{!\postDominates{$DG, C, d.getBasicBlock()$}} {
		        $\textit{PostDominatesFlag} = false$\;
		        $BC = BC \setminus \{C\}$\;
                % break\;
            }
		} \label{alg:ln:ft_end}
        \textcolor{blue}{\tcc{C3: Check for definition-free paths from $C$ to $v_{ex}$ using constant propagation}}
		\If{$\textit{PostDominatesFlag}$}{ \label{alg:ln:constant_propagation_start}
            \ForEach{$var \in C_\host$}{
                \If{!\isConstant{$G, var, C, v_{ex}$}}{
                    $BC = BC \setminus \{C\}$\;
                }
            }
        } \label{alg:ln:constant_propagation_end}
    }
    $\neck$ = \closestReachableFromEntry{$BC$}\; \label{alg:ln:ClosestToEntry}
    \uIf{$\neck = \emptyset$} { \label{alg:ln:null_st} %lor |\neck|~> 1
        \KwRet $null$\; \label{alg:ln:null_end}
    }\lElse{
        \KwRet $\neck$\; \label{alg:ln:return_neck}
    }
\end{algorithm}

%% file: evaluation.tex
\section{Evaluation}
\label{sec:evaluation}

Our experiments were designed to answer three questions:

\begin{itemize}
    \item \textbf{RQ1:} Can \name identify the \neck location correctly for command-line and configuration-file programs? (\S\ref{subsec:RQ1} \textit{shows that \name can identify the presence/absence of the \neck for $87.5\%$
    of the programs in the field study 
    and for $85.7\%$ of the programs in
    a previously unseen dataset}.)
    \item \textbf{RQ2:} How expensive is \name in terms of running time and memory usage? (\S\ref{subsec:RQ2} \textit{reports that total analysis time on $23$ apps is $8.5$ minutes, with max. memory consumption of $4.4$GB}.)
    \item \textbf{RQ3:} Can \name alleviate the manual efforts required to use existing debloating tools? (\S\ref{subsec:RQ3} \textit{demonstrates that {\name}'s integration with two debloating tools eliminates a user-annotation requirement without breaking their functionality}.)

\end{itemize}

\noindent

\noindent The evaluations were carried out on an Ubuntu $16.04$ PC with an Intel i7-5600U CPU @ 2.6GHz and $16$ GB RAM.

\begin{table}[tb!]
\centering
\caption{
Results of \name's evaluation. 
\machanged{Columns $3$ \& $4$ represent the number of pointers
% \twr{Clarify: do you mean ``\# of pointer variables declared'' or ``\# of statements that use a pointer variable''?}
and allocation sites, respectively.}
Column $5$ indicates whether \name could identify the \neck location correctly.
(``\no~(\yes)'' means that \name did not succeed ``out of the box,'' but was successful when provided with suitable stubs for two library functions.)
Column $6$ specifies \name's average running time in seconds, and column $7$ indicates the maximum amount of memory usage in MB (both over 10 runs).
}
\scalebox{0.7}{
\begin{tabular}{|c|r|r|r|c|r|r|}
\hline
\textbf{Program} & \textbf{kLOC} & \textbf{\#ptr}	& \textbf{\#alloc} &\textbf{\shortstack{Correct\\ \neck}} & \textbf{\shortstack{Analysis\\ Time}} & \textbf{Memory} \\ \hline
\rowcolor{gray!40}\multicolumn{7}{|c|}{\textbf{End-user Programs}} \\ \hline
curl-7.47.0  & 31.6 & 10228	& 785 & \no & 119.6  & 1526  \\ \hline
date-8.32      & 56.9  & 7613 &	979         & \yes  & 26.9       & 698   \\ \hline
diff-2.8         & 37.2  & 5331 &	842         & \yes  & 2.9        & 213   \\ \hline
du-8.32        & 109.1  & 19168	& 2756       & \yes  & 3.5       & 299   \\ \hline
echo-8.32      & 11.0  & 1687 &	305         & \yes  & 0.5        & 72    \\ \hline
gzip-1.2.4       & 26.7  & 2952	& 420         & \yes  & 0.6        & 112   \\ \hline
id-8.32        & 15.1  & 2368 &	418         & \yes  & 1.8        & 135   \\ \hline
kill-8.32      & 12.1 & 1819 &	348         & \yes  & 1.7        & 130   \\ \hline
objdump-2.33     & 1049.0 & 209651	&21984       & \yes  & 37.7       & 1261  \\ \hline
psql-15     & 189.1 & 37921 &	3798       & \yes  & 7.4        & 305   \\ \hline
readelf-2.33     & 413.7 & 76842 &	8325        & \yes  & 4.5        & 400   \\ \hline 
sort-8.32      & 55.9  & 9902 &	1549         & \yes  & 4.0         & 254   \\ \hline
tcpdump-4.10.0   & 608.6 & 152999	 &11945         & \yes  & 24.1       & 1431  \\ \hline
uniq-8.32      & 14.8  & 2340 &	442         & \yes  & 1.9        & 136   \\ \hline
wc-8.32        & 17.8 & 3000 &	509          & \yes  & 1.8        & 138   \\ \hline
wget-1.17.1      & 165.5 & 30069 &	4036        & \yes  & 2.7        & 276   \\ \hline

% chown-8.32$^*$ & 3.37 &5168	&717 & \yes  & 1.8    & 172  \\ \hline
% rm-8.32$^*$ & 3.4 & 5282 &	728   &  \yes & 1.25    & 173   \\ \hline

\rowcolor{gray!40}\multicolumn{7}{|c|}{\textbf{Server Programs}}               \\ \hline
bind-9.15.8      & 1755.1  & 326408	&41391      & \yes  & 228.8      & 4292  \\ \hline
DNSProxy-1.17    & 3.4  & 583 & 79          & \yes  & 0.1        & 53    \\ \hline
httpd-2.4.51     & 179.0  & 75032 &	6927       & \no~(\yes) & 2.2        & 278   \\ \hline
knockd-0.5       & 10.9 & 2062	& 177          & \yes  & 0.1        & 57    \\ \hline
lighttpd-1.4.54  & 174.0 & 34932 &	3527       & \yes  & 1.7        & 195   \\ \hline
mini-httpd1-1.19 & 16.4  & 2935	& 323          & \yes  & 1.9        & 147   \\ \hline
Nginx-1.19.0     & 589.2  & 116710 &	9307       & \yes  & 32.7       & 1232    \\ \hline  
PostgreSQL-15 & 4626.3 & 880507 &	126401 &  \no  & 575.6 &  8313.7 \\ \hline

% Memcached$^*$  & 73.1 & 22167 &	2691   & \yes  & 0.4   & 126    \\ \hline  
% Redis$^*$  & 1437.36 & 302314 &	65001   & \no  & XX   & XX    \\ \hline 
% Squid$^*$  & XX   &  &&  & XX   & XX    \\ \hline 
% AFL-fuzz$^*$  & 164.9 & 26548 &  3181 &  \yes  & 1.9   & 253.4    \\ \hline
\end{tabular}
}
\label{tab:evaluation:results}
\end{table}

\subsection{\rqOne: Accuracy of \name}
\label{subsec:RQ1}

The accuracy of \name is defined as its ability to determine whether the subject program contains an acceptable
single-element \neck.
A correct answer means that \name identified one of the suitable \neck locations, or correctly indicated that the program lacks any suitable location.

We conducted our evaluation using two sets of programs:
(i) the programs considered in the manual field study (\S\ref{sec:manualStudy}, Table~\ref{tab:motivation:prelimStudy}), and
(ii) the $21$ programs listed in Table~\ref{tab:evaluation:unseen},
which were neither involved in the manual field study nor examined to determine \neck properties.
The latter set was introduced to provide an unbiased test of accuracy results.

\input{unseen_dataset}

\subsubsection{Accuracy based on Manual-Field-Study Dataset}
For each program in the evaluation dataset, we measured the accuracy of \name in identifying the correct \neck location using the following methodology:

\begin{enumerate}
    \item Manually annotate the target program's source code with the
    (single-element) \neck location that is closest to the entry point of \texttt{main} in terms of distance along control-flow-graph edges,
    and generate the LLVM IR bitcode.
    This information serves as ground truth. 
    \item
    Pass an un-annotated LLVM IR bitcode of the same target program to \name, which annotates one basic block of the bitcode as the \neck. 
    \item Check whether the basic block identified by \name matches the ground truth. If the check passes, then \name is successful in identifying the \neck.
\end{enumerate}

\noindent

\name accurately reports the existence/absence of an acceptable \neck for $21$ out of $24$ programs ($87.5\%$)
in Table~\ref{tab:evaluation:results}.
\name fails to report an accurate \neck location for \texttt{curl}, \texttt{httpd}, and \texttt{PostgreSQL} because of the following reasons:
 
\begin{itemize}
    \item
    \emph{A Swiss-army-knife program}:
    as mentioned in \S\ref{sec:manualStudy}, \texttt{PostgreSQL} requires a multi-element \neck.
    While this category is not within the purview of the \name, we still ran \name on \texttt{PostgreSQL}.
    \name was able to correctly identify the entire set of elements of the multi-element \neck as \neck candidates, but returned the one nearest to the program entry point as per line~\ref{alg:ln:ClosestToEntry} of Alg.~\ref{alg:boundary-algorithm}.
    % Additionally, the remaining \necks do exist within the \neck candidates list as suggested by Alg.~\ref{alg:boundary-algorithm}.
    % If PostgreSQL is excluded from consideration, \name's accuracy improves from $87.5\%$ to $91.3\%$.
    \item
    \emph{Definitions of two argument-parsing functions are unavailable}:
    % \name on \texttt{httpd} fails as it uses a library for parsing the configuration arguments that hinders \name's taint analysis from correctly propagating the configuration hosting variables.  
    \texttt{httpd} uses \texttt{libapr (Apache Portable Runtime)} (specifically \texttt{apr_getopt_init} and \texttt{apr_getopt}) to parse command-line arguments.
    Only the declarations of these functions exist in the LLVM bitcode;
    their definitions are not available, which prevents \name from performing taint analysis---and thus from identifying the correct \neck location.
    However, when provided with suitable stubs---i.e., taint-analysis summaries that describe the dependencies of outputs on inputs in \texttt{apr_getopt_init} and \texttt{apr_getopt}---\name is able to identify the \neck location correctly.

    \item
    \emph{
    Configuration logic and main-computation phase in the \textbf{same} procedure called from \texttt{main}}:
    \name always places the \neck inside \texttt{main}, just after the call site that contains the code identified as the configuration logic.
    However, in \texttt{curl} both the configuration logic and the main-computation logic reside in the \textbf{same} callee of \texttt{main}.
    \name is unable to identify the \neck correctly because no location in \texttt{main} separates the configuration logic from the main-computation logic.
    
\subsubsection{Accuracy based on Previously Unseen Programs}
\label{sec:AccuracyOnPreviouslUnseen}
\machanged{This dataset contains programs that were not employed to deduce the \neck's traits (\S\ref{sec:manualStudy}).
}

\noindent
\textbf{Selection of Subject Programs.}
This dataset was acquired by cloning starred C/C++ projects from GitHub with $3k$ stars or more, which yielded $100$ repositories. %C=50 and C++=50
% \todo{EB: 76 or not?}
We then excluded repositories that
(a) include other programming languages (such as Python, JavaScript, etc.),
(b) incorporate GUI functionality,
(c) did not contain an entry point (i.e., firmware), or
(d) did not build successfully.
This process yielded the $21$ repositories listed in Table~\ref{tab:evaluation:unseen}.
Of these repositories, $18$ possess a single-element \neck, one has a multi-element \neck, and two lack any \neck (see Table \ref{tab:neckCategories}).
These findings corroborate the outcomes of the field study (\S\ref{sec:manualStudy}).

\machanged{For each program in the unseen dataset, we assessed the accuracy of \name %in pinpointing a suitable \neck location 
using the following approach:
\begin{enumerate}
    \item Generate the LLVM IR bitcode (without any instrumentation) and analyze it using \name.
    % \item Analyze LLVM IR bitcode using \name.
    \item Manually examine the source code to compare with \name's outcome.
    \name's result is considered correct if (i) the \neck identified by \name aligns with the manually identified location, or (ii) \name does not detect any \neck and the manual inspection confirms the absence of an
    acceptable single-element \neck (i.e., the program has multi-element \necks or no \neck).
\end{enumerate}
}

\begin{comment}
\begin{table}[tb!]
\centering
\caption{
Results of \name's evaluation based on the unseen programs. See Table~\ref{tab:evaluation:results} caption for column details.
}\vspace{-0.1cm}
\scalebox{0.65}{
\begin{tabular}{|c|r|r|r|c|r|r|}
\hline
\textbf{Program} & \textbf{kLOC} & \textbf{\#ptr}	& \textbf{\#alloc} &\textbf{\shortstack{Correct\\ \neck}} & \textbf{\shortstack{Analysis\\ Time}} & \textbf{Memory} \\ \hline
% \rowcolor{gray!40}\multicolumn{7}{|c|}{\textbf{End-user Programs}} \\ \hline
AFL-fuzz  & 164.9 & 26548 &  3181 &  \yes  & 1.9   & 253.4    \\ \hline
chown-8.32 & 3.37 &5168	&717 & \yes  & 1.8    & 172  \\ \hline
% \rowcolor{gray!40}\multicolumn{7}{|c|}{\textbf{Server Programs}}               \\ \hline
Memcached  & 73.1 & 22167 &	2691   & \yes  & 0.4   & 126    \\ \hline  
Redis  & 1437.36 & 302314 &	65001   & \no  & XX   & XX    \\ \hline 
rm-8.32 & 3.4 & 5282 &	728   &  \yes & 1.25    & 173   \\ \hline
Squid  & XX   &  &&  & XX   & XX    \\ \hline
\end{tabular}
}
\label{tab:evaluation:unseen}
\end{table}
\end{comment}

Table~\ref{tab:evaluation:unseen} presents the \neck-identification results for the unseen programs.
% For $18$ out of the $21$ 
For $18$ out of the $21$ programs, \name accurately reports the existence/absence of an acceptable (single-element) \neck.

The three cases for which \name reported an inaccurate result were as follows:
 \item
    \emph{rethinkdb}: \name returns a single location in \texttt{main}, rather than a set of locations that constitute a multi-element \neck.
  \item
    \emph{masscan}: Although the identified \neck satisfies all \neck properties, the code before this \neck does not actually parse the configurations.
    Instead, it configures the program to report debug information in case it crashes.
    A single variable, \texttt{is_backtrace}, is initialized inside a loop that parses \texttt{argv}, whereas the rest of the configuration-hosting variables are parsed inside another loop inside the function \texttt{masscan_command_line}, which is called from \texttt{main}.
    \name does mark the latter location as a candidate \neck initially, but it is then eliminated because it is not the closest to the entry of \texttt{main} (line~\ref{alg:ln:ClosestToEntry} of Alg.~\ref{alg:boundary-algorithm}).

  \item
    \emph{Redis}:
    Despite running on a server with 192GB of memory,
    % specifically to analyze \texttt{Redis},
    \name's data-flow analysis phase exhausted memory.

\end{itemize}

% \noindent We verified that \name handles correctly plausible scenarios listed above and is able to identify the \neck location accurately, except \texttt{PostgreSQL} case, where \textbf{Multiple distinctive \neck locations} exist.

\subsection{RQ2: Performance of \name}
\label{subsec:RQ2}

\begin{figure}[tb!]
\centering
  \begin{tabular}{cc}
    \includegraphics[width=.45\linewidth]{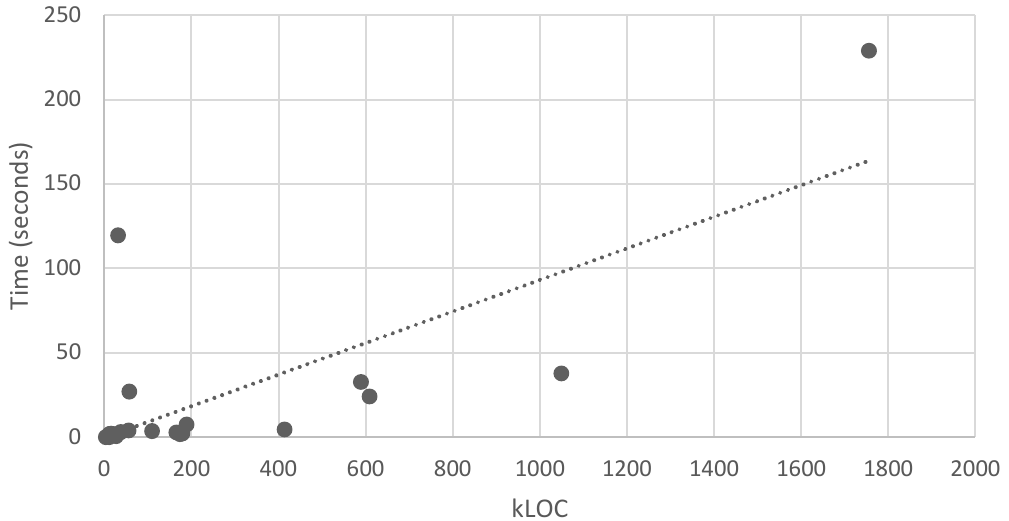}
    &
    \includegraphics[width=.45\linewidth]{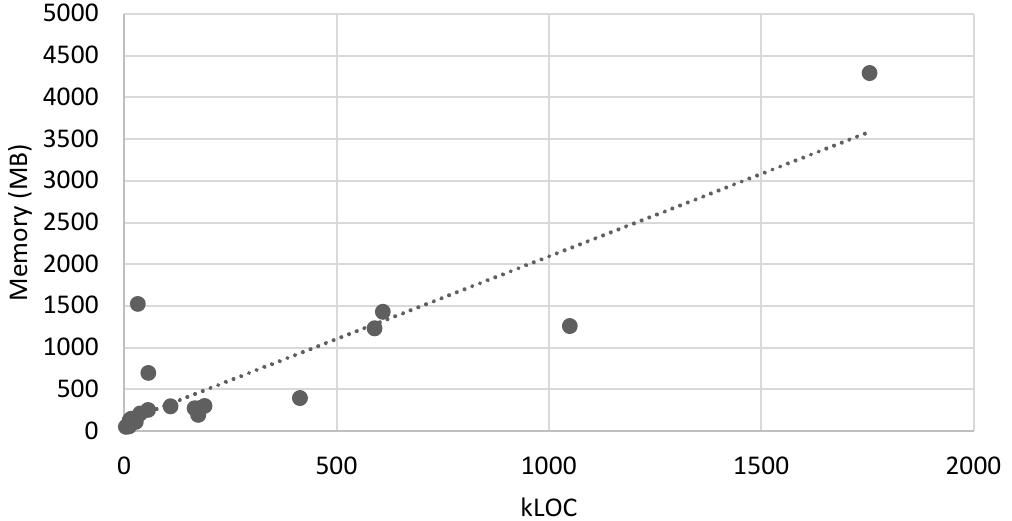}
  \end{tabular}
  \caption{Lines of code versus time (left) and memory (right).}
  \label{fig:ScatterPlots}
\end{figure}

We measured the analysis time and memory usage for each program in Table~\ref{tab:evaluation:results} (averaged over 10 runs) using the UNIX \texttt{time} tool, which provides the total analysis time and the peak memory usage that a process uses. As shown in Fig.\ \ref{fig:ScatterPlots}, analysis times and memory consumption scale roughly linearly with lines of code.
(In both plots, the outliers are \texttt{curl} and \texttt{bind} on the high side and \texttt{readelf} and \texttt{objdump} on the low side.)

\name's taint analysis is influenced by a program's characteristics, such as the number of pointer variables (both \# of pointer variables declared and \# of statements that use a pointer variable),
stores/loads, indirect call sites, etc.~\cite{10.1145/3315568.3329965, 9610731}, which affect the number of data-flow facts that need to be propagated through the target program and are not strictly linked to the number of lines of code.
Table~\ref{tab:evaluation:results} provides information about the number of pointers and allocation sites in each program.
We calculated the Pearson Correlation Coefficient to establish the strength of the relationship between kLOC, number of pointers, allocation sites, analysis time, and memory usage. The Correlation Coefficient in general was over $0.95$, which indicates
% as shown in Table~\ref{tab:correlation}. 
a positive linear relationship between these factors.

% \vspace{-0.25cm}
\subsection{\rqThree: Effectiveness of \name}
\label{subsec:RQ3}
We discuss three case studies to demonstrate the benefits of integrating \name in \sota software-debloating tools~\cite{LMCAS,Temporal} and configuration-error detection tools~\cite{PCHECK}. 
% These tools rely on the presence of a \neck between the configuration phase to the main computation phase. 
The case studies explain how the integration of \name with these tools alleviates the manual effort required by a developer, thus making the tools easier to use. 
These tasks involve combing through source code to track \texttt{argc} and \texttt{argv} usage, and potentially debugging complex programs to ensure the \neck is reachable from the program's entry point.

%Further, we integrate \name as a part of the \lmcas \neck mining step to see it's effectiveness to aide \lmcas in performing software debloating. \lmcas assumes the \neck mining step to be done manually by a developer and highlights that the step can be performed within $5-10$ minutes. In all practical scenarios, this assumption is based on the fact that the developer has a thorough understanding of program's internal structure. Therefore, the time required to perform the manual \neck mining step heavily relies on the skill and knowledge of the developer without any upper bound. In our evaluation, we show that \name can be used as a part of \lmcas to debloat programs (see Table \ref{tab:dataset}) in order to offset the manual work required by a developer thereby reducing the burden on a developer. With \name, the developer does not need to know the internal structure of a program at all.

\subsubsection{Integration with \lmcas}
\lmcas~\cite{LMCAS} is a debloating tool that applies partial evaluation to specialize a program to a particular run-time configuration.
% Its debloating pipeline relies on knowing the \neck between the configuration phase and the main-computation phase.
Currently, the \lmcas user is required to annotate the program to specify the \neck.
% \lmcas provides a semi-automated utility called neck miner to identify the \neck. However, neck miner relies on tracking configurations using def-use chain, which is imprecise in contrast to the taint analysis applied by \name. This utility entirely depends on the user as they need to inspect a set of control-flow properties manually.
% For real-world programs, manual identification can be burdensome and error-prone.

We incorporated \name into the \lmcas pipeline:
a program's LLVM IR bitcode is annotated by \name with the program's \neck location, and then passed to \lmcas.
We evaluated our extension of \lmcas on the programs \machanged{(obtained from the \lmcas dataset)} listed in Table~\ref{tab:lmcas}.
Our aim was to understand the efficiency gained by integrating \name into \lmcas.
Improved efficiency means (1) reducing or eliminating the burden on the \lmcas user of identifying a program's \neck location, and (2) making sure that automatic \neck identification does not affect \lmcas's ability to create a correctly working debloated program.
Correct functionality can be validated by running the debloated programs with the supplied test inputs, omitting the flags specifying the features for which they have been debloated.
We matched the output of the debloated program with that of the original program,
which was supplied the appropriate feature flags and the same inputs.
If the output is the same, the debloated program is considered to have preserved the functionality.
We also check that the \name-annotated programs do not crash the \lmcas debloating pipeline. 

Table~\ref{tab:lmcas} reports the results of this experiment. \name reduces the analysis time of the \neck-identification step in \lmcas from minutes to a few seconds.
It also eliminates human error due to manual analysis. Finally, \name+ \lmcas preserves the functionality of  debloated programs (and \name does not break the debloating pipeline of \lmcas).

% To measure the efficiency of the pipeline, we measure the memory usage and the running time of each stage in the pipeline => \neck + Partial Interpretation + Constant Conversion +  MultiStage Simplifications + Collect Statistical info. 

% In addition, we also measure the running time, lines of code and memory usage of the run. Since these metrics are unique for each run of the program, we ran each experiment 100 times and computed the average over all the runs.

% \nam{we actually evaluated for all programs in table 1} \ma{have we debloated curl, knockd, and nginx?} \nam{i remember for curl, neck tool didnt work on my PC. but for knockd it did - the only issue was i could not compile knockd and test it}

% For the purpose of our evaluation, we selected two programs, a server program (i.e., \texttt{mini-http}) and client program (i.e., \texttt{wget}) that are supported by \lmcas. 

% We identify \neck in these programs using \name and provide the annotated programs to \lmcas. We then debloated these programs based on the settings described in~\cite[Table 5]{LMCAS}. We observe: (1) the supplied programs do not break the debloating pipeline, and (2) the functionality is preserved in the debloated programs. 

\begin{table}[tb!]
\centering
\caption{
Effectiveness of \name in facilitating \neck-identification for \lmcas.
}
\begin{threeparttable}  % Start the threeparttable environment here
\begin{tabular}{|c|r|r|c|c|}
\hline
\multirow{3}{*}{Program} & \multicolumn{2}{c|}{\neck ident. time} & \multirow{3}{*}{\shortstack{Correct\\ \neck}} & \multirow{3}{*}{\shortstack{Functionality\\ Preserved}} \\
\cline{2-3}  
   & \shortstack{LMCAS \\(minutes)$^a$} & \shortstack{\name\\ (seconds)} &   &    \\\hline
chown & 5 - 10 & 0.6  & \yes & \yes \\ \hline 
date  & 5 - 10 & 1.8 & \yes & \yes \\ \hline 
gzip  & 5 - 10 & 0.3 & \yes & \yes \\ \hline 
rm    & 5 - 10 & 0.7 & \yes & \yes \\ \hline 
sort  & 5 - 10 & 1.4 & \yes & \yes  \\ \hline 
uniq  & 5 - 10 & 0.4 & \yes & \yes   \\ \hline           
\end{tabular}
\begin{tablenotes}  % Use tablenotes for footnotes in tables
\item[a] Reported by the LMCAS authors~\cite{LMCAS}.
\end{tablenotes}
\end{threeparttable}  % End the threeparttable environment
\label{tab:lmcas}
\end{table}

% This demonstrates \name entirely eliminates the need of semi-automated utility, neck miner. 
% Accordingly, \name improves the use of \lmcas to users who do not need to have an in-depth knowledge about the intended program to be debloated.

% \ma{TODO: i may need to add a table shows debloating settings for the selected 2 programs or reference table 6 in \lmcas paper}

% We also observed that \name is practical, and has low to moderate footprint in terms of running time and memory for any downstream analysis.
%We also show that \name can be used as a part of \lmcas to debloat chosen programs (see Table \ref{tab:dataset}).

% \naman{todo: review}
%By measuring the running time of the \name for each program, we argue that using \name inside a downstream task such as \lmcas, we are able to reduce the manual effort required by a developer who is expected to have in-depth knowledge about the program. 

% program the analysis time for the heuristic analysis was 2
% seconds on average, and it took 5−10 minutes to perform
% the manual part of structural analysis. This amount of time
% is acceptable, given that \neck identification is performed
% only once per program

% Our programs, the raw as well as the processed data produced in our evaluation is available in our artifact~\cite{Artifacts}.

\subsubsection{Integration with Temporal Specialization}
The \texttt{temporal-specialization} tool~\cite{Temporal} reduces the attack surface of a server program by disabling unneeded system calls.
The system calls to disable are determined by splitting server programs into phases of initialization and serving:
any system call never used in the serving phase is to be disabled once the (manually identified) ``transition point'' between the phases is reached.
Clearly, a transition point is the same concept as a \neck, so \name can be applied to the problem of transition-point identification.
The functions for the initialization and serving phases are called the \emph{master} and \emph{worker}, respectively. 
% But these two functions are identified manually.
Our evaluation tested whether the function calls representing each phase are correctly separated.
We performed the following steps:
(i) for each program in Table~\ref{tab:temporal} (obtained from the \texttt{temporal-specialization} dataset),
we ran \name to identify the \neck;
(ii) as ground truth, we used the master and worker functions employed in the \texttt{temporal-specialization} evaluation:
If the configuration logic identified by \name includes a call to the master function, and the main-computation logic identified by \name contains a call to the worker function, we considered \name to have identified an appropriate transition point.
As shown in Table~\ref{tab:temporal}, \name identified an appropriate transition point in each example, \machanged{except for \texttt{Redis}, where the analysis again ran out of memory (OOM).}

\begin{table}[tb!]
\centering
\caption{
Effectiveness of \name in facilitating transition-point identification for the \texttt{temporal}\texttt{-specialization} tool.}
\begin{tabular}{|c|c|c|c|}
\hline
Program   & \shortstack{Correct\\ \neck} & \shortstack{Master func. in \\ config. logic} &\shortstack{Worker func. in \\ main logic} \\ \hline
Bind       & \yes & \yes & \yes \\ \hline
Memcached  & \yes & \yes & \yes \\ \hline
Nginx      & \yes & \yes & \yes \\  \hline     
Apache     & \yes & \yes & \yes \\  \hline
Lighttpd   & \yes & \yes & \yes \\  \hline
Redis      & OOM & OOM & OOM \\  \hline
% redis info based on redis-server.ll
\end{tabular}
\label{tab:temporal}
\end{table}

\subsubsection{PCHECK}
PCHECK is an analysis tool that aims to detect configuration errors.
It generates configuration-checking code to be invoked after program initialization.
PCHECK requires users to identify the \neck manually.

We could not integrate \name with PCHECK because its implementation is unavailable.
PCHECK's dataset includes three Java and three C/C++ programs, but the only \necks defined in the paper are 
for \texttt{Squid} (C++) %\twr{or is Squid C++?} \ma{it's C++}
and \texttt{HDFS} (Java).
Thus, we focused our attention on Squid (786k LOC).
\name successfully identified \texttt{Squid}'s \neck in 41 seconds.

%% file: unseen_dataset.tex
\begin{table*}[tb!]
\centering
\caption{
\machanged{Results of \name's evaluation based on the previously unseen programs. 
% Columns 2 and 3 provide information about the repository category and 
Popularity is based on the number of stars. 
Column 5 reports the \neck type based on the manual inspection of the source code.
Column 6 indicates the outcome of \name's analysis: (P) means a \neck is present;
(A) (for ``absent'') indicates that no \neck is present. 
% \twr{I don't think it is right to say that \name obtained the correct answer for \texttt{redis}. By design, \name returns a single location, and it hoists that location to \texttt{main}. Thus, \name cannot return a multi-element-\neck answer. If I were a referee, I could imagine rejecting the paper on the basis of this claim. Better to claim that \name got the correct answer for $17$ programs than risk rejection.}
% Column 7 denotes if \name's findings are aligned with the manual inspection.}
}}
\scalebox{0.6}{
\begin{tabular}{|c|c|r|r|c|c|c|}
\hline
\textbf{Repo (Program)} & \textbf{Category} &\textbf{Popularity} &\textbf{kLOC} & \neck Type & \textbf{\name Results} &\textbf{Accuracy}\\ \hline
AFLplusplus(afl-fuzz) & Software Testing & 3.7k & 164.9 & Single & P & \yes \\ \hline
blurhash(blurhash_encoder) & Image Processing & 13.8k & 58.6 & Blended & A & \yes\\ \hline
Caffe(caffe) & Machine Learning & 33.5k& 17.2 & Blended & A & \yes\\ \hline
fish-shell(fish) & Utility & 21.8k& 696.3 & Single & P & \yes\\ \hline
coreutils(chown) & Utility & 3.5k & 3.37 & Single & P & \yes\\ \hline
coreutils(rm) & Utility & 3.5k & 3.4& Single & P & \yes\\ \hline
GoAccess(goaccess) & Web & 16.4k & 189.7 & Single & P & \yes \\ \hline
hashcat(hashcat) & Utility & 17.4k& 978 & Single & P & \yes \\ \hline
jq(jq) & Utility & 25.3k & 300.1 & Single & P & \yes \\ \hline
masscan(masscan) & network & 21.3k & 223 & Single & P & \no \\ \hline
memcached(memcached) & Data & 12.6k & 73.1 & Single & P & \yes\\ \hline 
$n^3$(nnn)  & Data & 16.4k & 46.4 & Single & P & \yes\\ \hline 
Redis(redis) & Data & 60k & 1437.4 & Single & OOM & OOM \\ \hline 
rethinkdb(rethinkdb) & Data & 26.2k & 839.1 & Multi-element & P & \no \\ \hline 
skynet(skynet) & Games & 12k &629.7 & Blended & A & \yes \\ \hline
% (Squid)  & XX   & & &1.6k\\ \hline
Tesseract(tesseract) & Image Processing & 51.9k & 2204.6 & Single & P & \yes\\ \hline
the_silver_searcher(ag) & Utility & 25k& 28.2 & Single & P & \yes \\ \hline
tmux(tmux) & Utility &29.5k & 593.6 & Single & P & \yes\\ \hline
trojan(trojan) & Network & 17.8k & 346.6 & Single & P & \yes\\ \hline
twemproxy(nutcracker) & Network &11.8k & 152.8 & Single & P & \yes\\ \hline
wrk(wrk)  & Web & 34.5k & 1203.3 & Single & P & \yes\\ \hline
\end{tabular}
}
\label{tab:evaluation:unseen}
\end{table*}

%% file: threatsToValidity.tex
%\vspace{-0.25cm}
\section{Threats to Validity} \label{sec:threats}

We outline threats to the validity of our approach, along with the applied mitigations: 

\begin{itemize}
  \item \textbf{Scope of the study}
    (Internal \& external validity).
    We investigated programs whose configurations are provided through \cli or configuration files.
    Some of our findings may not generalize to other kinds of software, such as event-driven programs (e.g., Android programs).
    For the field study, we selected a diverse set of widely used, mature programs.
    However, to avoid bias, we evaluated \name using popular programs
    from GitHub that we had \emph{not} used to identify \neck properties.

  \item \textbf{Robustness of \neck properties} (External validity).
     The properties used by \name to infer \neck locations were inferred from the 24 programs in Table\ \ref{tab:motivation:prelimStudy}. 
     % which represent a number of different kinds of software systems.
     Moreover, the properties used by \name do not depend on heuristics like function/variable names and data types like int/string.
     \name also does not exploit special idioms that are used by some programmers for parsing programs' configurations.
     For instance, we observed that the GNU Coreutils programs use a particular idiom (i.e., the invocation of the function \texttt{getopt_long} inside a while-loop) for parsing command-line parameters.
     Instead, we decided to have \name rely on high-level structural properties
     that are driven by program-configuration semantics. 
     \machanged{The evaluation of $21$ unseen programs validates the identified \neck properties in common programs and confirms \name's effectiveness in boundary identification.}
     \item \textbf{Incorrect propagation of data flows} (External Validity).
     Our taint analysis, discussed in \S\ref{sec:bc}, is sound under practical assumptions, such as system and \texttt{llibC} calls behaving as expected. It does not account for \texttt{lsetjmp} and \texttt{llongjmp} usage or dynamically loaded code via \texttt{dlopen/dlsym}. If these assumptions are violated, the analysis becomes unsound.
\end{itemize}

\noindent
Finally, there is the question of the soundness \name's results when run on the three different kinds of \neck cases.
\begin{itemize}
  \item
    \textbf{Single-element}: \texttt{masscan} shows that \name is fallible, and can return an incorrect answer when a single-element \neck exists (see \S\ref{sec:AccuracyOnPreviouslUnseen}).
    The \texttt{masscan} result constitutes both a false positive and a false negative.
  \item 
    \textbf{Multi-element}: in our limited experience, \name identifies the locations of a multi-element \neck as (individual) candidates, but because \name returns just a singleton location in \texttt{main}, the answer returned is a false positive.
  \item 
    \textbf{Blended}:
    \name returns \texttt{null}, because the ``blended'' case involves violations of single-element-\neck properties.
\end{itemize}

%% file: relatedWork.tex
%\vspace{-0.3cm}
\section{Related Work} \label{sec:rw}

% \eb{This section appears a tad short, I would suggest to extend it a bit, trying to save space in other parts, particularly the middle part (methodology).} \ma{we've commented out some stuff, I'll make sure putting them back if space allows}

% \twr{The min-cut material is a candidate to cut. Also, the point of a related-work section is not to just enumerate other work, but to show how that work compares to the work described in the paper.}

\noindent\textbf{Multi-cut for Program Decomposition.} 
Multi-cut algorithms~\cite{multicut} have been used in several program-optimization methods~\cite{Speculation,loop-decomposition,Vertex-Cut}. 
% Johnson et al.~\cite{Speculation} used min-cut to decompose a program to maximize program parallelism---creating new threads in sequential programs.
Ma et al.~\cite{Vertex-Cut} presented a vertex-cut framework on LLVM IR graphs to partition coarse-grained dataflow graphs into parallel clusters to improve
% reduce data communication and to improve the scalability and 
performance of applications in multi-core systems.
% Ootsu et al.~\cite{loop-decomposition} developed a min-cut method to enhance the performance of loops. Their method applies loop unrolling to the loop during the decomposition process by identifying a min-cut set in the weighted control-flow graph of a program.
In our work, only a degenerate form of min-cut is used:
the algorithm identifies the set of articulation points, each member of which constitutes a cut-set of size $1$.
However, \name's static taint analysis is an improvement on the data-dependency analysis used in \cite{Vertex-Cut}, which relies on dynamically generated traces.

\noindent\textbf{Tracing Program Configurations.}
% TRIMMER~\cite{trimmer_journal} tracks \texttt{argv} to annotate variables and memory objects that may hold parameters read from configuration files and/or program inputs. 
% Rabkin et al.~\cite{configurationAPIs} leverage static analysis to automatically infer a program’s configuration for improving a program's documentation.
% They focus on identifying key-value-style configurations by tracing the arguments of certain configuration APIs in the constructed call-graph.
% It traces certain configuration APIs, which are used for reading a program's configuration. Their approach finds the earliest configuration processing point in a call chain. Similarly,
LOTRACK~\cite{Lotrack_TSE,Lotrack_ASE} applies taint analysis for identifying all code that is influenced by load-time configurations in Android and Java programs. In Android applications, the identification of a \neck appears to be less of a problem: because 
% Android apps are essentially plugins in the Android framework, with a dedicated lifecycle, all configuration logic has typically already been completed by the time the app starts to execute its main activity. 
Android apps, essentially plugins with a specific lifecycle in the Android framework, usually have their configuration logic completed (i.e., typically inside \texttt{onCreate} and before \texttt{onStart}) by the time of executing their main activity.
% \twr{Not sure what ``main activity'' refers to. Is that a technical term in the Android framework? If so, I would say that the lifecycle supported includes ``XXX'' and a ``main activity.'' Thus, it is typical that configuration logic executes in XXX and the main-computation logic executes as the main activity.} \ma{my understanding ``activity'' has a lifecycle (callback APIs e.g.,[onCreat, onStart,..]). While ``main activity'' typically refers to the entry point of the application where the user starts interacting with the app, and is specified inside the manifest file. The above lifecycle applicable to the main activity. So my hunch is that Eric refers to onCreate and before onStart(), where the configurations are initialized. Hence, somehow the initial config is confined inside the onCreate}
Hence, this program point can thus serve as the \neck. This observation does not hold for regular Java programs, then we foresee \name can be leveraged to solve this \neck identification challenge in this context.
Finally, LOTRACK relies on the assumption that configuration APIs are known; however, identifying such APIs can be cumbersome. \name does not require configuration APIs, the taint analysis of \texttt{argv} is sufficient to identify configuration-hosting variables including APIs that read configuration files. %Thus, \name taint analysis can support these tools.  

%% file: conclusion.tex
%\vspace{-0.25cm}
\section{Conclusions and Future Work}
\label{sec:Conclusion}

This paper presents an algorithm and tool, called \name, to statically identify programs' configuration logic. 
Our evaluation on widely used C/C++ command-line and configuration-file programs confirmed the existence of a \neck and found that
\name \emph{automatically} identified a suitable \neck for $87.5\%$ of the programs.
Finally, we demonstrated an application of \name to reduce the manual-annotation burden in software-debloating and error-detection tools.

\begin{comment}
This paper presents an algorithm and tool, called \name, to decompose programs based on the observation that programs can often be split into configuration logic and main logic. 
We conducted a preliminary study, which confirmed that such a split was possible in $23$ of $24$ programs in our corpus.
We then formalized the problem as one of finding a set of configuration-hosting variables, an articulation point in the control-flow graph, and properties of the configuration-hosting variables that hold (i) at the articulation point, and (ii) along all paths from the articulation point to the exit of the program. We evaluated \name
on widely used C/C++ command-line and configuration-file programs, and found that a program's \neck can be determined in an \emph{automated} manner for $91\%$ of $23$ programs.
Finally, we demonstrated an application of \name to reduce the manual-annotation burden in two software-debloating tools. 
\end{comment}

% \name can be integrated with other tools to reduce the burden of manual analysis. %(\S\ref{subsec:RQ3}).
Additionally, we envision that \name can be used as a linting tool to alert developers that they have intertwined a program's configuration logic with its main-computation logic.
Thus, \name supports ongoing initiatives~\cite{ACCORD} to promote configurability as a first-class programming concept.

In future work, we would like to examine the existence of \necks in GUI programs and event-driven programs. Furthermore, multi-cut algorithms could allow \name to handle Swiss-Army-knife cases.
% Finally, \name should have additional interesting applications, for example, in facilitating program documentation.

% This allows one to integrate \name with other applications such as program debloating that previously suffered from not being able to automatically compute this piece of information.
% \ps{@Mohannad: How has the world improved now? We can automatically compute a program's \neck and use this information to help with program comprehension and other applications such as automated program debloating.}
% \eb{This is currently a pretty standard section that simply rehashes what we did. I prefer it when authors really present here their main \emph{conclusions}. What worked well, what did not? What are future directions?}
% \twr{I would be happy to axe this section because it adds nothing to what is already available in the Abstract. Like Eric, I am not a fan of a Conclusion section that does not add additional insight about the work. Many referees do not agree, but if the choice is between the Conclusion or Fig.~\ref{fig:ScatterPlots}, I'd take Fig.~\ref{fig:ScatterPlots} any day. }